\documentclass[review]{elsarticle}
\usepackage{amsmath,amsfonts,amssymb}
\usepackage{graphics}
\usepackage{graphicx}
\usepackage{setspace}
\usepackage{tocloft}
\usepackage[linesnumbered,ruled]{algorithm2e}
\usepackage[noend]{algpseudocode}
\usepackage{lineno,hyperref}
\usepackage[caption=false,font=footnotesize]{subfig}
\modulolinenumbers[5]

\journal{Journal of Optics \& Laser Technology}

\begin{document}

\begin{frontmatter}

\title{Temporal Analog Optical Computing using an On-Chip Fully Reconfigurable Photonic Signal Processor}
\author[a]{Hossein Babashah}
\author[a]{Zahra Kavehvash\corref{mycorrespondingauthor}}
\author[a]{Amin Khavasi}
\author[b]{Somayyeh Koohi}
\address[a]{Sharif University of Technology, Department of Computer Engineering, Azadi Ave., Tehran, IRAN}
\address[b]{Sharif University of Technology, Department of Electrical Engineering, Azadi Ave., Tehran, IRAN}

\begin{abstract}
This paper introduces the concept of on-chip temporal optical computing, based on dispersive Fourier transform and suitably designed modulation module, to perform mathematical operations of interest, such as differentiation, integration, or convolution in time domain. The desired mathematical operation is performed as signal propagates through a fully reconfigurable on-chip photonic signal processor. Although a few number of photonic temporal signal processors have been introduced recently, they are usually bulky or they suffer from limited reconfigurability which is of great importance to implement large-scale general-purpose photonic signal processors. To address these limitations, this paper demonstrates a fully reconfigurable photonic integrated signal processing system. As the key point, the reconfigurability is achieved by taking advantages of dispersive Fourier transformation, linearly chirp modulation using four wave mixing, and applying the desired arbitrary transfer function through a {cascaded} Mach-Zehnder modulator {and phase modulator}. Our demonstration reveals an operation time of $200~ps$ with high resolution of $300~fs$. To have an on-chip photonic signal processor, a broadband photonic crystal waveguide with an extremely large group-velocity dispersion of $2.81 \times {10^{6}}~\frac{{{ps^2}}}{km}$ is utilized. Numerical simulations of the proposed structure reveal a great potential for chip-scale fully reconfigurable all-optical signal processing through a bandwidth of $400~GHz$.
\end{abstract}

\begin{keyword}
Integrated Optics, Integrability, Optoelectronics, Nonlinear Optics, Dispersion, Fourier Optics.
\end{keyword}

\end{frontmatter}

\section{Introduction}
It is more convenient for an integrated photonic structure to perform any mathematical operations, such as convolution with an arbitrary function and pulse shaping, in time domain~\cite{1,2,3,4,5,6,7}. Therefore, analog optical computing has gained widespread applications in optical communication and real-time spectroscopy for processing optical signals in time domain.\par
One of the most important features in digital signal processing (DSP) is the processing speed, mostly restricted by the electronic sampling rate. In an optical network, signal processing is carried out by DSP, which is responsible for electronic sampling, as well as optical-to-electrical (OE) and electrical-to-optical (EO) conversions. As an approach to achieve power-efficient and high-speed signal processing capability in an optical network, we can implement signal processing unit directly in the optical domain using a photonic signal processor to avoid the need for electronic sampling and OE and EO conversions~\cite{13,14,15}.\par
So far, numerous photonic signal processors have been proposed based on either discrete components or photonic integrated circuits~\cite{13,14,15,16,7,18,19,20,21,22,azana1,azana2}. Photonic signal processors based on discrete components, such as fiber-bragg-grating~\cite{3,azana1,azana2}, usually have good programming abilities but are bulkier and less power efficient, whereas a photonic integrated signal processor has a much smaller footprint and a higher power efficiency. A photonic signal processor can be used to implement various types of applications, such as optical pulse shaping~\cite{53}, arbitrary waveform generation~\cite{13}, temporal integration~\cite{20}, temporal differentiation~\cite{21}, optical dispersion compensation~\cite{19}, and Hilbert transformation ~\cite{23}. These functions are basically the building blocks of a general-purpose signal processor performing signal generation and fast computing. Fast computing processes, such as temporal integration, temporal differentiation, and convolution facilitate important applications~\cite{23,24,25,26,27,28,29,30,31,32,33,34}. For example, a photonic integrator, as a device that able to perform time integral of an optical signal, has a key role in dark soliton generation~\cite{24}, optical memory~\cite{25}, and optical analog-digital conversion~\cite{26}. Moreover, a photonic temporal differentiator~\cite{29}, as a device performing temporal differentiation of an optical signal, is employed for all-optical Fourier transform~\cite{30,31}, temporal pulse characterization~\cite{32}, and the demultiplexing of an optical time division multiplexed (OTDM) signal~\cite{33}, for example. Although the photonic implementations of these functions have been reported so far~\cite{20,21,22,27,29}, an optical signal processor is usually designed to perform a specific function with no or very limited reconfigurability. For general-purpose signal processing, however, a photonic signal processor is desired to perform multiple functions with high reconfigurability~\cite{1}. In~\cite{1}, the reconfigurability is achieved by controlling the injection currents to the active components (i.e. semiconductor-optical-amplifiers) of the signal processor which still yields specific functionality such as differentiation, integration, and Hilbert transform. Consequently,  applying any new frequency transfer function or spatial transformation requires a new complicated design of the structure. Therefore, implementing arbitrary transfer function in time domain is worth to be noticed. Pulse shaping has been performed before by transferring the signal from time domain to spatial domain and applying appropriate filters in spatial domain~\cite{7}. Nevertheless, as the main drawback of these structures, they necessitate precise optical alignment procedure for fiber optic applications, in which light must be coupled with low loss out of and back into single-mode optical fiber.\par
To address above limitations, in this paper, a general time-domain analog optical computing structure is proposed which can be implemented in an integrated photonic system. The proposed structure which is of a very small size, provides the possibility of highly compact, potentially integrable architectures within much smaller volumes and, in some cases, over sub-wavelength length, ensuring controlled manipulation and processing of the incoming signal. As a novel idea, we proposed a structure that takes advantages of dispersive Fourier transformation to implement any arbitrary transfer function by a simple time-modulation in the frequency domain. In other words, our work is the time domain counterpart to the idea proposed in~\cite{51,52,josab} which performs analog optical computing in spatial domain using lenses and metasurfaces.\par
Dispersive Fourier transformation has thus far been utilized for separating the frequency components in time domain in order to overcome the slow response of the detector~\cite{8} or low speed of the electro-optic phase modulator used for pulse shaping~\cite{azana1,azana2}. Nevertheless, no analog optical computing, i.e., computational Fourier transformation, has been performed based on a dispersive Fourier transform structure. Still, the use of dispersive Fourier transformation in analog optical computing requires specific considerations. To make the structure implementable in an integrated photonic system, we implemented the dispersive Fourier transformation~\cite{8,9,10} with an on-chip structure with high group-delay-dispersion (GDD) benefiting from time lens for chirp modulation \cite{foster2008silicon}. Moreover, time-domain modulation for multiplying the signal and arbitrary transfer function in frequency domain can be performed through  a cascaded Mach-Zehnder modulator and optical modulator~\cite{11,111}. \par
The paper is organized as follows:  A review on dispersive Fourier transformation and time lens is discussed in section 2. Section 3 present the proposed temporal analog optical computing structure. Simulation results are discussed in Section4; Section 5 concludes the paper.\par
\section{A Review on Dispersive Fourier Transformation and Time Lens}

Time-domain Fourier transformation has been implemented through passing light via dispersive media~\cite{9}. Dispersive Fourier transformation (DFT) maps the broadband spectrum of a conventionally ultrashort optical pulse into a time stretched waveform with its intensity profile mirroring the spectrum using chromatic dispersion. It is known that a dispersive element can be modeled as a linear time-invariant (LTI) system with a transfer function given by $H(\omega ) = \left| {H(\omega )} \right|{e^{j\varphi (\omega )}}$, where $| {H(\omega )}|$ and $\varphi (\omega)$ are the magnitude and phase response of the dispersive element at angular frequency $\omega$, respectively. Mathematically, the phase response $\varphi (\omega)$ can be expanded in Taylor series. This dispersive element is conventionally a long length single mode fiber~\cite{8}. The propagation of an optical pulse through a dielectric element with up to the second-order dispersion coefficients (assuming negligible higher-order dispersion coefficients within the bandwidth of interest) can be described with the following transfer function and impulse response:
\begin{equation}
\label{eq1}
\begin{array}{l}
H(\omega) = \left| {H(\omega)} \right|{e^{j{\varphi _0}}}{e^{j{{\dot \varphi }_0}\omega}}{e^{j\frac{1}{2}{{\ddot \varphi }_0}{\omega^2}}}\mathop  \to \limits^{{\Im ^{ - 1}}} \\
h(t) = {e^{j{\varphi _0}}}{e^{ - j\frac{1}{{2{{\ddot \varphi }_0}}}{{(t - {{\dot \varphi }_0})}^2}}}
\end{array}
\end{equation}
where $| {H(\omega )|}$ is engineered to be constant or have weak dependence to the angular frequency and $\varphi (\omega)$ equals to $\beta (\omega) L$ where $\beta (\omega)$ and $L$ are propagation constant and waveguide's length, respectively. The stretched pulse can be therefore approximated by~\cite{12}:
\begin{equation}
\label{eq2}
y(t) = x(t) * h(t) = \int\limits_{ - \infty }^\infty  {x(\tau ){e^{j{\beta _0 L}}}{e^{ - j\frac{1}{{2{{\ddot \beta }_0 L}}}{{(t - \tau  - {{\dot \beta }_0 L})}^2}}}d\tau } 
\end{equation}
setting ${\tau _R} = \tau  + {\dot \beta _0 L}$ we have:
\begin{equation}
\label{eq3}
\begin{array}{l}
y(t) = {e^{j{\beta _0 L}}}\int\limits_{ - \infty }^\infty  {x({\tau _R} - {{\dot \beta }_0 L}){e^{ - j\frac{1}{{2{{\ddot \beta }_0 L}}}{{(t - {\tau _R})}^2}}}d{\tau _R}}  = \\
{e^{j{\beta _0 L}}}{e^{ - j\frac{1}{{2{{\ddot \beta }_0 L}}}{t^2}}}\int\limits_{ - \infty }^\infty  {x({\tau _R} - {{\dot \beta }_0 L}){e^{ - j\frac{1}{{2{{\ddot \beta }_0 L}}}{\tau _R}^2}}{e^{ + j\frac{1}{{{{\ddot \beta }_0 L}}}t{\tau _R}}}d{\tau _R}} 
\end{array}
\end{equation}
to compensate for the term ${e^{j\frac{1}{{2{{\ddot \beta }_0 L}}}{\tau _R}^2}}$, the input signal, $x(t)$, is modulated with a quadratic phase modulation as follows:
\begin{equation}
\label{eq4}
x(\tau ) = {x_0}(\tau ){e^{ + ja{{(\tau  + {{\dot \beta }_0}L)}^2}}}
\end{equation}
where $a$ is the chirp factor. By considering $a={\frac{1}{{2{{\ddot \beta }_0}L}}}$ we yield the following equation for $y(t)$:
\begin{equation}
\label{eq5}
{ \begin{array}{l}
	\tilde y(t) = {e^{j({\beta _0 L} - \frac{1}{{2{{\ddot \beta }_0 L}}}{t^2})}}\int\limits_{ - \infty }^\infty  {{x_0}({\tau _R} - {{\dot \beta }_0 L}){e^{j\frac{1}{{{{\ddot \beta }_0 L}}}t{\tau _R}}}d{\tau _R}} \\
	= {e^{j({\beta _0 L} + \frac{{t + {{\dot \beta }_0 L}}}{{{{\ddot \beta }_0 L}}}{{\dot \beta }_0 L} - \frac{1}{{2{{\ddot \beta }_0 L}}}{t^2})}}{\left. {{X_0}(\omega)} \right|_{\omega = \frac{{t + {{\dot \beta }_0 L}}}{{ - {{\ddot \beta }_0 L}}}}}
	\end{array}}
\end{equation}
where $X(\omega ) = \Im \{ x(t)\} $, is the Fourier transform of the input pulse. 
This phase modulation is significantly larger than the $10\pi$ possible
phase shift using an electro-optical phase modulator, and therefore an alternative scheme can be realized using a parametric nonlinear wave mixing process such as four wave mixing \cite{foster2008silicon}. In this approach, a Gaussian pump pulse propagates through a dispersive medium with a GVD equals to ${\ddot \beta}_p$ and length of $L_p$ that is much longer than the dispersion length of the pulse. As a result, the pulse undergoes temporal broadening and is linearly chirped. The broadend linearly chirped signal which is proportional to ${e^{j\frac{{{\tau ^2}}}{{2{{\ddot \beta }_p}{L_p}}}}}$ is then used in the four wave mixing process as the pump pulse. The output signal denoted as idler in the four wave mixing can be filtered and it is calculated as follows:
\begin{equation}
\label{eq6}
{E_{idler}(t)} = {E^2}_{pump}(t){E^*}_{input}(t)
\end{equation}
where $E_{idler}(t)$, $E_{pump}(t)$, and  $E_{input}(t)$ are the output, pump and input  electric field amplitude in the four wave mixing process. The output electric field frequency, ${\omega _{idler}} = 2{\omega _{pump}} - {\omega _{input}}$, is obviously different from the input and the pump frequency. This frequency could be separated in the output through using a bandpass Brag filter.
It should be here mentioned that, the chirp phase modulation could be avoided by making the term ${\tau _R}^2/2{\ddot \beta _0 L} <  < 1$ ,  for all values of ${\tau _R}$    between 0 and pulse duration, $T$.  This is possible if and only if a long length of the dispersive medium and/or a large group velocity dispersion (GVD), ${\ddot \beta _0}$ is used.

\begin{figure}
	\centering
	\includegraphics[width=1.1\linewidth]{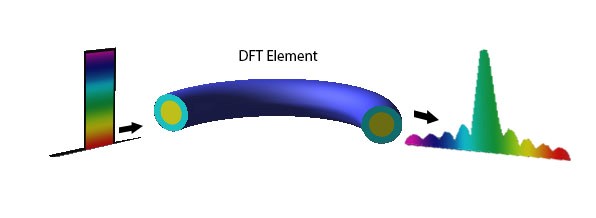}
	\caption{Schematic structure of dispersive Fourier transformation for an optical signal as it passes through a waveguide with up to second order dispersion coefficient.}
	\label{fig:fig1}
\end{figure}

To summarize, it is clearly noticeable from Eq. (\ref{eq5}) that the output temporal pulse envelope is proportional to the Fourier transform of the input pulse multiplied by a phase factor. In this way, DFT or dispersion-induced frequency-to-time mapping could be obtained through a dispersive media. This process is schematically shown in Fig. \ref{fig:fig1}.
\begin{figure}
	\centering
	\includegraphics[width=1.1\linewidth]{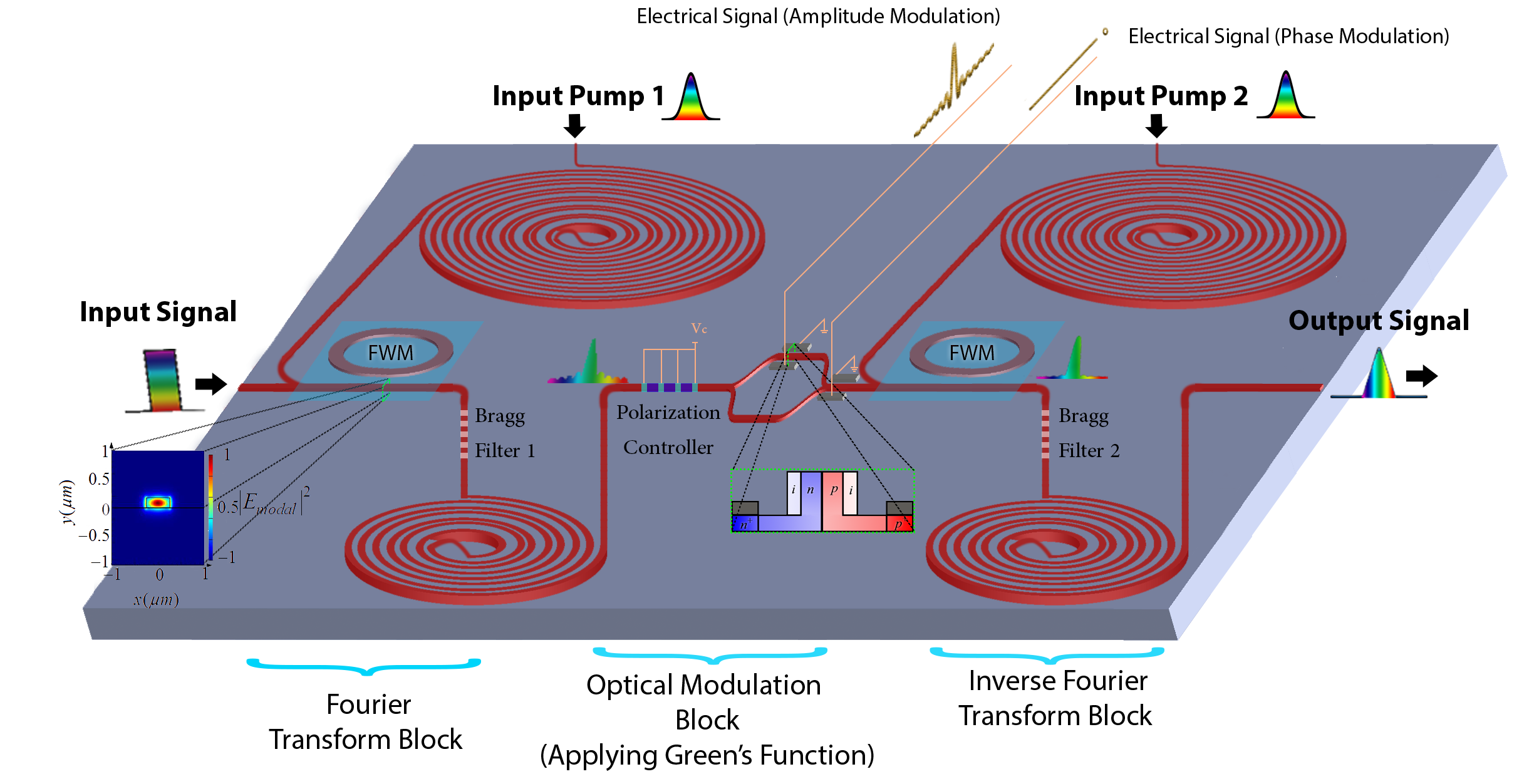}
	\caption{Schematic of the proposed temporal analog optical computing structure, including four dispersive Hydex waveguides \cite{Hydex} and a cascaded MZI modulator and phase modulator based on p-n diode inside a waveguide. The first block performs Fourier transform by utilizing a linear GVD and a time lens, while the second block represents a modulator and multiplies the desired kernel, and finally the third block performs inverse Fourier transformation using a negative slope linear GVD with a time lens.}
	\label{fig:fig2}
\end{figure}
\section{Proposed Temporal Analog Optical Computing Structure}
In the proposed temporal analog optical computing structure, we have utilized Green's function method to implement arbitrary time-domain operations such as integration, differentiation, and convolution. Considering the input time-domain signal as $y(t)$, the output signal, $o(t)$, of the system with Green's function $g(t)$ is computed as follows:
\begin{equation}
\label{eq7}
o(t) = y(t) * g(t) = \int {y(\tau )g(t - \tau )} d\tau 
\end{equation}
which is the well-known convolution operation. This operation can be written in Fourier domain as a simple multiplication:
\begin{equation}
\label{eq8}
{\tilde o({f_t}) = \tilde y({f_t})\tilde g({f_t})}
\end{equation}
where, tilde indicates the Fourier transform of signals and ${f_t}$  represents the temporal frequency variable. Based on this basic mathematical relation, the proposed temporal analog optical computing structure mainly consists of three building blocks, as shown in Fig. \ref{fig:fig2}. The first building block is a temporal Fourier transform structure which performs Fourier transformation on the input signal. The resultant signal in the Fourier domain is then passed through a temporal optical modulator which multiplies the input signal with the desired Green's function in the Fourier domain. Given that the input signal and the Green's function are both in Fourier domain, their multiplication is equivalent to their convolution in Fourier domain. And finally, the third block is responsible for performing inverse-Fourier-transformation ($IFT$) on the modulator output to yield the result of convolution with desired Green's function. Inverse Fourier transformation requires the conjugate kernel of $FT$ operator and thus, according to Eq. (\ref{eq5}), we need a media with negative GVD, ${\ddot \beta _0}$,  for performing $IFT$. It is worth to mention that $FT$ or $IFT$ of a signal can be used instead with considering the fact that $IFT\{ s(t)\}  = \tilde s( - f_t)$. The $FT$ and $IFT$ blocks, as well as the optical modulator will be designed and analyzed in the following subsections.

\subsection{Recipe for temporal FT and IFT}
\begin{figure}
	\centering
	\includegraphics[width=1.1\linewidth]{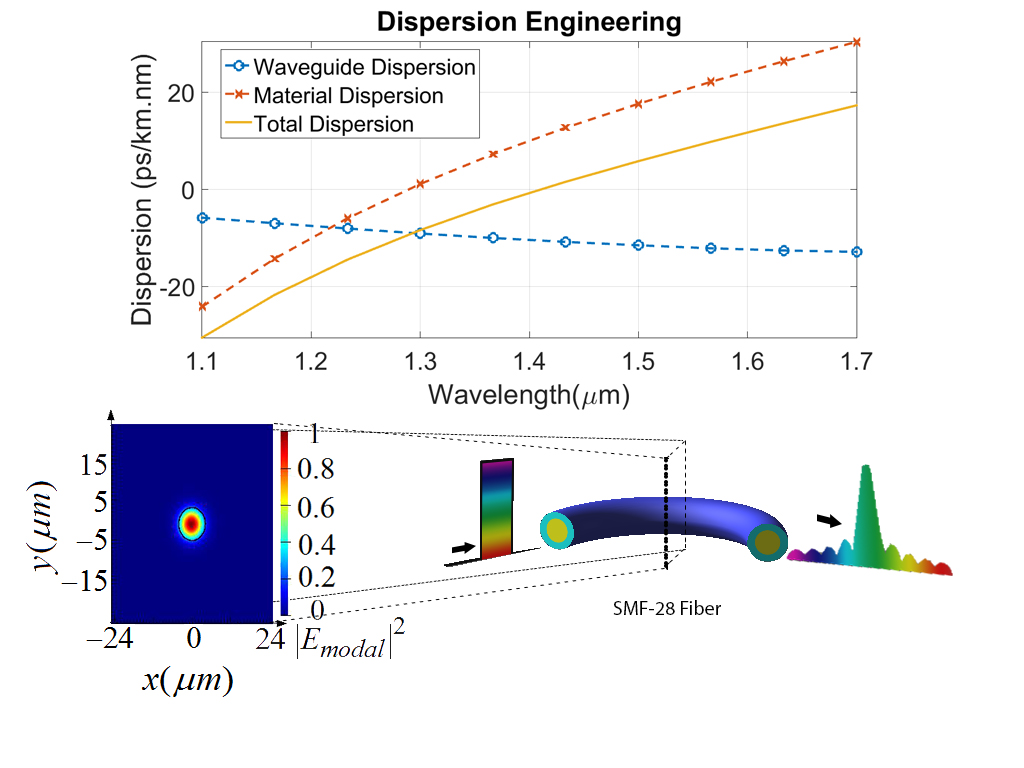}
	\caption{Engineering the waveguide group-velocity dispersion. Dispersion of a structure is obtained as the sum of its material dispersion and waveguide dispersion. The waveguide dispersion can be engineered by adjusting the effective mode area of the waveguide to have a constant GVD in the desired bandwidth for DFT.}
	\label{fig:fig3}
\end{figure}

Considering the DFT structure discussed in Section 2, we require a constant or linear second order dispersion coefficient. In other words, we should engineer the waveguide’s GVD through tailoring the dimensions of the waveguide cross section. The GVD for a given waveguide cross section can be approximated as the sum of material’s GVD and its waveguide’s GVD (Fig. \ref{fig:fig3}). While the material’s GVD is typically normal (${\ddot \beta _0} > 0$) at short wavelengths and anomalous (${\ddot \beta _0} < 0$) at longer wavelengths, the waveguide’s GVD exhibits the opposite behavior. Thus, by adjusting the waveguide cross section, the waveguide dispersion can be tailored to engineer the total dispersion, as shown in Fig. \ref{fig:fig3}.
This waveguide can be a single mode fiber~\cite{8}, linearly chirped fiber-brag-grating (FBG)~\cite{12}, silicon nitride waveguide~\cite{45}, or photonic crystal waveguide~\cite{46}. All these waveguides have the capability of coupling an optical signal passing through an optical communication line to and/or from them through tapered couplers~\cite{37,38}. Furthermore, they can have linear or constant GDD in the optical communication spectrum. Constant GDD, ${\ddot \phi _0} = {\ddot \beta _0}L = C'$, means that the group delay, $\Delta \tau  = {\dot \beta _0}L$  is a linear function of frequency:
\begin{equation}
\label{eq9}
\begin{array}{l}
\frac{{\partial {{\dot \beta }_0}}}{{\partial \omega }} = \ddot \beta  = C\\
\Rightarrow {{\dot \beta }_0} = \ddot \beta \Delta \omega \\
\Rightarrow \Delta \tau  = {{\dot \beta }_0}L = \left( {\ddot \beta L} \right)\Delta \omega 
\end{array}
\end{equation}
\begin{figure}
	\centering
	\includegraphics[width=1\linewidth]{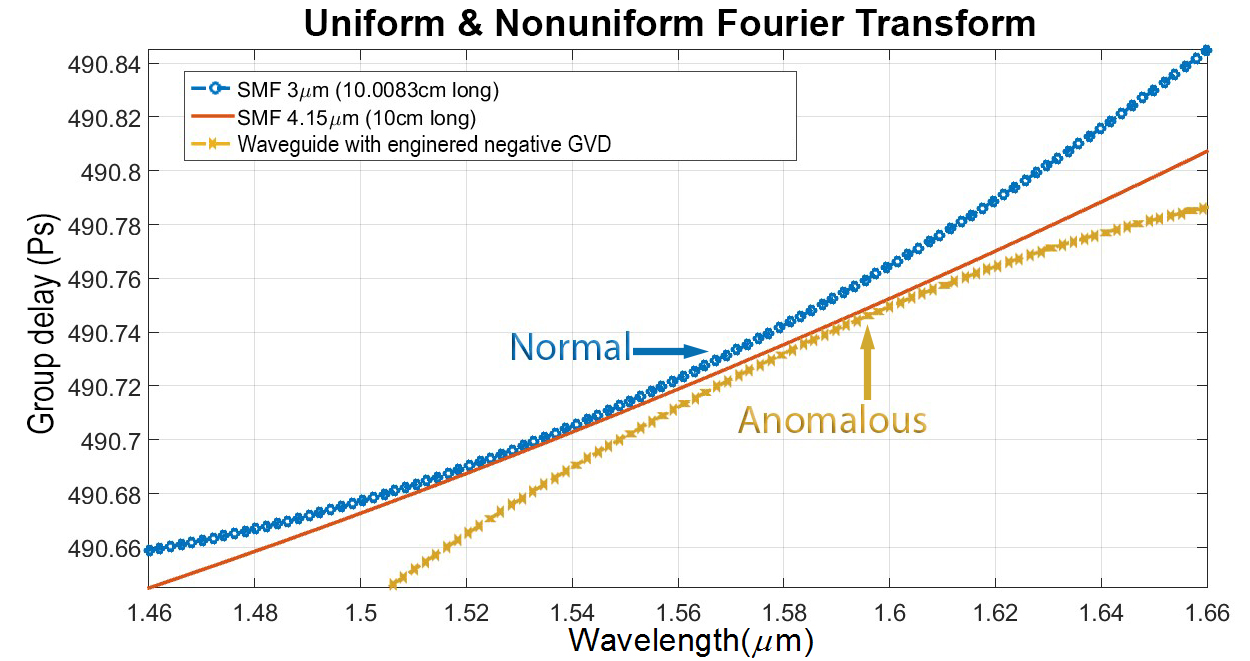}
	\caption{Frequency to time mapping for a single mode fiber and a waveguide with different core widths to engineer group delay and different lengths for easy comparison. Linear dispersion leads to uniform $FT$ while normal and anomalous dispersions lead to a nonuniform $FT$.}
	\label{fig:fig4}
\end{figure} 
\begin{figure}
	\centering
	\mbox{\subfloat[]{\label{subfig5:a} \includegraphics[height=4.6cm]{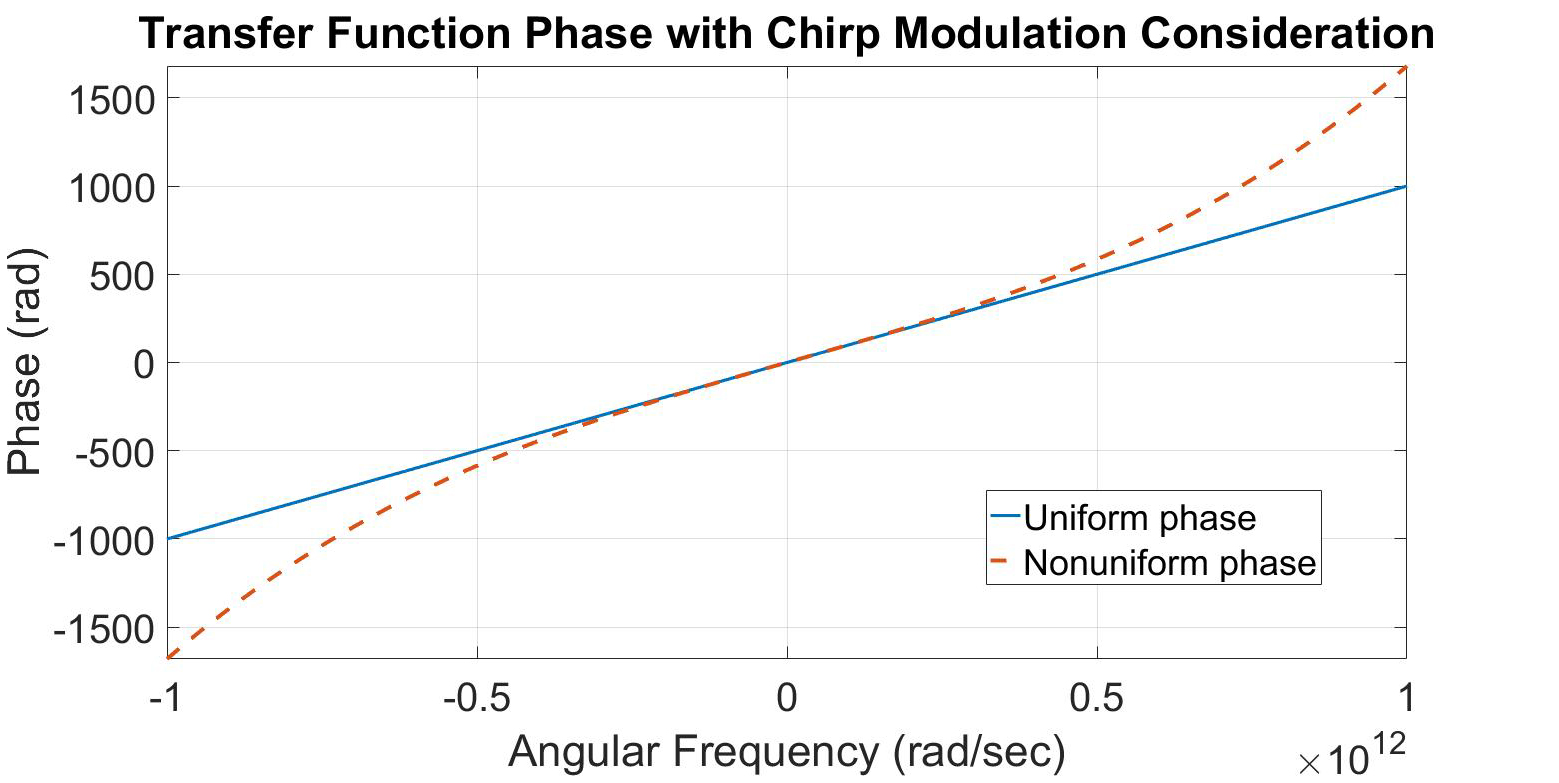}}}
	\mbox{\subfloat[]{\label{subfig5:b} \includegraphics[height=4.6cm]{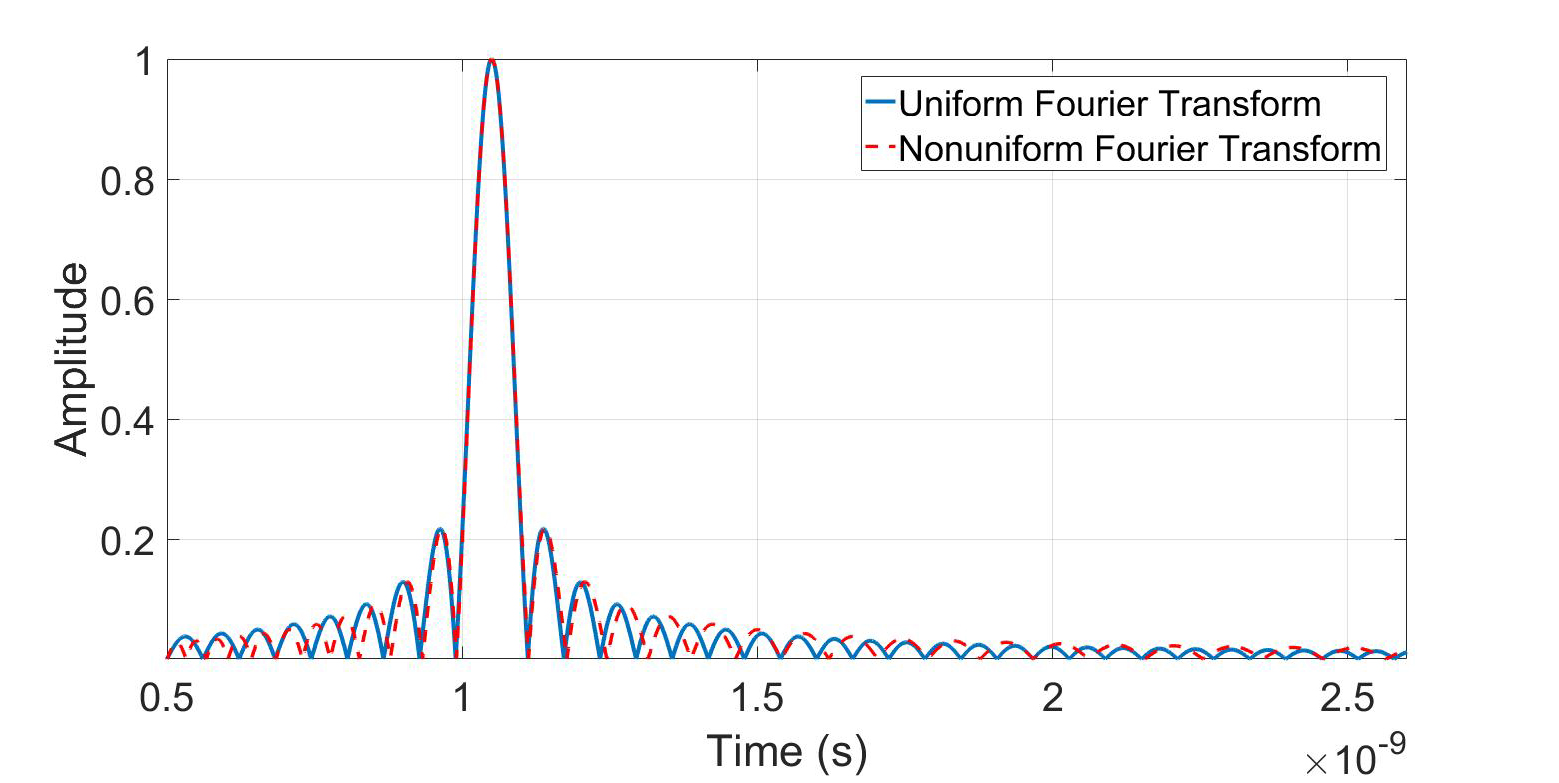}}}
	
	\caption{(a) Effective uniform and nonuniform applied phases to the input optical signal after linear chirp modulation with chirpyness of $\frac{1}{{{\beta _2}z}}$. (b) Non-uniform vs uniform $FT$ for a single square input pulse of 0.1ns width and dispersion parameters of ${\beta _1} = 1 \times {10^{3}}~\frac{ps}{km}$ , ${\beta _2} = 1.53 \times {10^{ 2}}\frac{{{ps^2}}}{km}$ and ${\beta _3} = 1.02 \times {10^{ 2}}~\frac{{{ps^3}}}{km}$ passed through a 40 kilometer long fiber.}
	\label{fig:fig5}		
\end{figure}
thus the time-frequency mapping is linear meaning that the resultant frequency samples are placed uniformly in the time domain which leads to uniform $FT$. Still, linear GDD, i.e. nonzero and constant ${{\phi _0}^{(3)}} = {{\beta_0} ^{(3)}}L = C'$, leads to nonlinear functionality of group delay versus frequency:
\begin{equation}
\label{eq10}
\begin{array}{l}
\frac{{{\partial ^2}{{\dot \beta }_0}}}{{\partial \omega }} = {\beta ^{(3)}} = C\\
\Rightarrow {{\dot \beta }_0} = {\beta ^{(3)}}\Delta {\omega ^2} + \ddot \beta \Delta \omega \\
\Rightarrow \Delta \tau  = {{\dot \beta }_0}L = \left( {{\beta ^{(3)}}L} \right)\Delta {\omega ^2} + \left( {\ddot \beta L} \right)\Delta \omega 
\end{array}
\end{equation}
\begin{figure}
	\centering
	\mbox{\subfloat[]{\label{subfig6:a} \includegraphics[height=4.3cm]{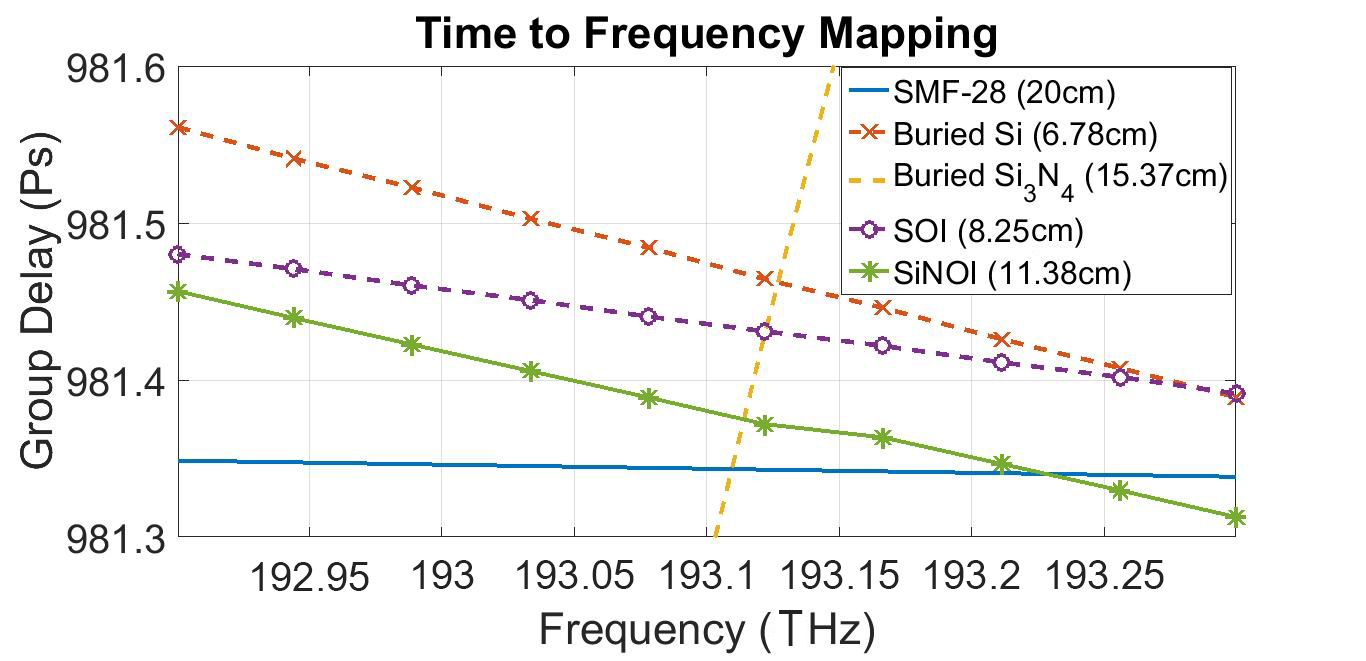}}}
	\mbox{\subfloat[]{\label{subfig6:b} \includegraphics[height=4.3cm]{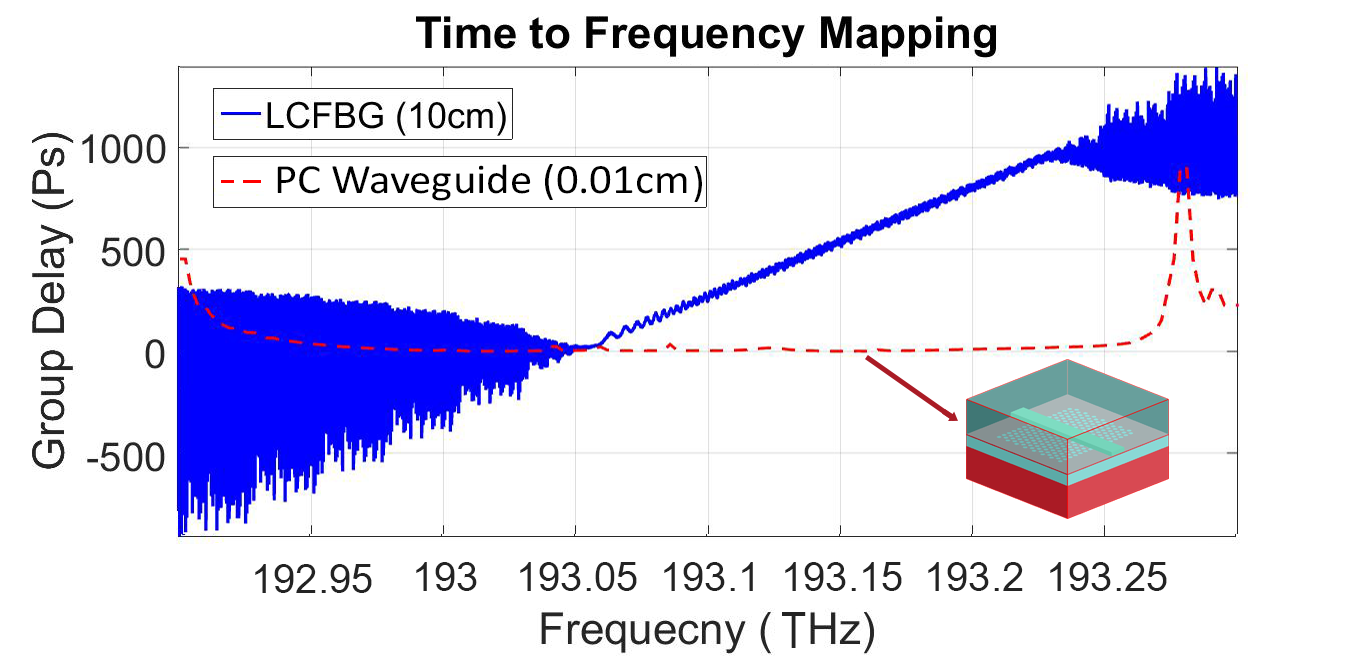}}}
	
	\caption{Group delay for different structures in communication frequency range. (a) Linear group delay for single mode fiber and different waveguide structures such as buried $Si$, buried $Si_3N_4$ in $SiO_2$, silicon on $SiO_2$, and $SiN$ on $SiO_2$ as insulator with waveguide cross section area of $0.5\times0.18 ~\mu m^2$, $0.5\times0.18 ~\mu m^2$, $1.15\times1.35 ~\mu m^2$ and $1.15\times1.35~ \mu m^2$, respectively. (b) Group delay for a LCFBG and a photonic crystal waveguide. The LCFBG consists of sine shape variation of refractive index and periodicity from 0.53433 to $0.53482~ \mu m$ and total chirp of $0.5~nm$ which clearly shows an approximately high constant GVD but includes ripples.  The main part of the PC waveguide is a low-loss $Si_{3}N_{4}$ rectangular core with $1 \mu m$ width and $400 nm$ thickness. Si/SiO2 photonic crystal layer underlying the core consists of two-dimensional triangular lattice of SiO2 pillars  with pitch of -400nm embedded in host Si layer. The diameter of SiO2 pillar is 250nm. Thin SOI layer (lOOnm thickness) serves as host medium for the PC layer, and buried oxide layer with $1 \mu m$ thickness in SOI wafer as bottom cladding \cite{ogawa2006broadband}. * Different waveguide length in parenthesis is selected for easy comparison.}
	\label{fig:fig6}		
\end{figure}
this nonunifrom time-frequency mapping results in a nonuniform Fourier transformation. The variation of group delay versus signal modulation wavelength for different values of ${{\beta_0} ^{(3)}}$ is shown in Fig. \ref{fig:fig4}. As depicted in this figure, the linear group-delay curve (frequency-time relation) i.e. zero  ${{\beta_0} ^{(3)}}$, maps the input time samples linearly to output frequency samples which results in uniform $FT$. In contrast, the non-linear group-delay curve, i.e. nonzero ${{\beta_0} ^{(3)}}$, transfers the uniform time samples to nonuniform output frequency samples. Still, nonuniform $FT$ is also acceptable as the dispersion curve is known for each structure because the Green's function and inverse $FT$ can also be performed with the same sampling pattern of the curve. An example of a nonuniform $FT$ as a result of linear GDD (nonzero ${{\beta_0} ^{(3)}}$) and constant GDD are shown in Fig. \ref{fig:fig5}. The group delay and GDD curves for different dispersive structures, including single mode fiber, linearly chirped fiber Bragg grating (LCFBG), Silicon or $Si_3N_4$ on insulator and buried $Si/SiO_2$ or $Si_3N_4/SiO_2$ waveguides, in the optical communication bandwidth, are shown in Fig. \ref{fig:fig6}\subref{subfig6:a} while the LCFBG and photonic crystal (PC) waveguide dispersive curve are plotted separately in Fig. \ref{fig:fig6}\subref{subfig6:b} due to their different scale. As shown in these figures, the group delay curves for these structures are approximately linear or have a linear differentiation. Thus, they are all appropriate for dispersive-Fourier transformation. The output Fourier-transformed signal is then multiplied by the desired transfer function by passing through the optical modulator which is explored in the next subsection. The resultant signal passes through an inverse Fourier transform ($IFT$) block which is also a DFT medium. However, given that the $IFT$ relation necessitates the complex exponential kernel in Eq. (\ref{eq5}) to be conjugated, the GVD, ${\ddot \beta _0}$, should be negative:

\begin{equation}
\label{eq11}
{\ddot \beta _0} =  - {\ddot \beta _0}'
\end{equation}

substitution of above equation in Eq. (\ref{eq5}) yields the following input-output relation:
\begin{equation}
\label{eq12}
{\small \begin{array}{l}
	s(t) = {e^{j({\beta _0 L} - \frac{1}{{2{{\ddot \beta L }_0}}}{t^2})}}\int\limits_{ - \infty }^\infty  {\tilde o({f_R} - {{\dot \beta }_0 L}){e^{-j\frac{1}{{\ddot \beta_0' L }}t{f_R}}}d{f_R}} \\
	= {e^{j({\beta _0 L} + \frac{{t - {{\dot \beta }_0 L}}}{{{{\ddot \beta }_0 L}}}{{\dot \beta }_0 L} - \frac{1}{{2{{\ddot \beta }_0 L}}}{t^2})}}{\left. {o(t')} \right|_{t' = \frac{{t + {{\dot \beta }_0 L}}}{{{{-\ddot \beta }_0' L}}}}}
	\end{array}}
\end{equation}
where $\tilde o(f_t)$  is the input to the $IFT$ block and $s(t)$ is the final output which is equal to its $IFT$ and thus based on Eq. (\ref{eq6}) is the convolution of the input signal and the desired Green's function that is injected to the optical modulator’s input.  The explanation of the used optical temporal modulator structure for performing multiplication and the analysis of the whole structure will be presented in the next subsections.

\subsection{Optical modulator}
\begin{figure}
	\centering
	\includegraphics[width=1\linewidth]{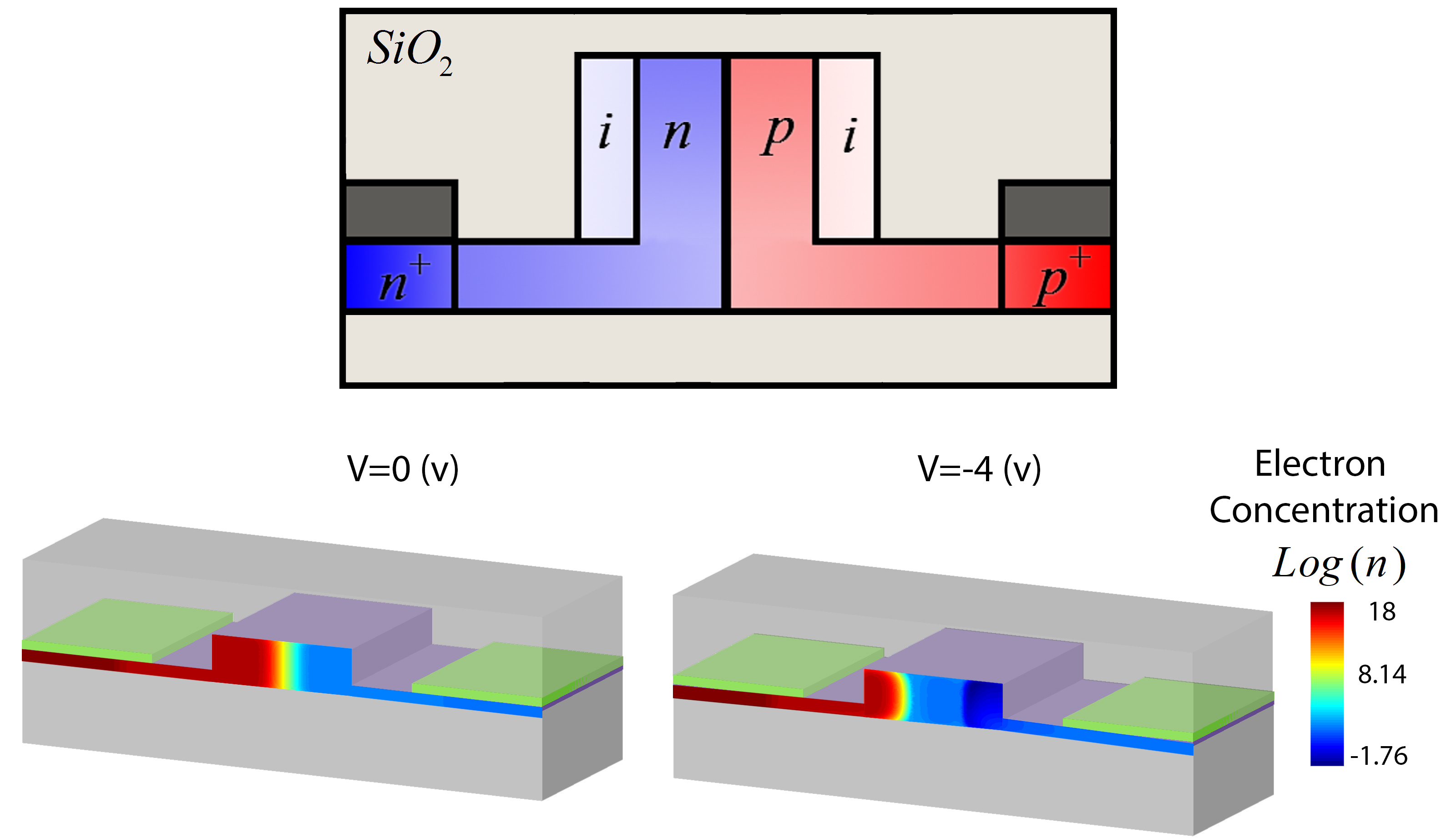}
	\caption{Schematic of the p-n junction implemented in an MZI Optical-modulator~\cite{44}. Increase in reverse bias of p-n junction leads to wider depletion region and refractive index increment.}
	\label{fig:fig7}
\end{figure}

Integrated optical modulators with high modulation speed, small footprint and large optical bandwidth are poised to be the enabling devices for on-chip optical interconnects~\cite{35,36}. Here, for optical modulation, a cascaded silicon Mach-Zehnder Interferometer (MZI) optical modulator is utilized~\cite{44}. MZI configuration is implemented using optical waveguides and Y-branches as beam splitters as shown in Fig. \ref{fig:fig2}. In these structures, the waveguides’ output signals will be interfered constructively or destructively upon having zero or $\pi$ radian phase shifts.  Using MZI configuration, an optical switch can be made by introducing a phase shift in the other waveguide as the voltage changes. In this structure, a p-n diode is constructed inside an optical waveguide as a phase shifter. The reverse bias on the p-n diode creates a depletion region which changes in size as a function of voltage as shown in Fig. \ref{fig:fig7}. The change in the number of carriers namely electron and holes in the waveguide introduces a small change in refractive index and accordingly lead to an optical phase shift and hence modulation. As this effect is very fast, p-n junction is used to construct a high-speed modulator by placing it inside an interferometer. Therefore, the modulator implemented by a MZI and a p-n junction on an arm of Y-branch is discussed here. Electrical signal is applied to p-n junction. Applying reverse voltages to the p-n junction leads to phase shifts between waveguides and ultimately the objective zero to $\pi$ phase shift between two different paths is achieved. Continuous laser light is fed as the input and the modulation signal is provided electrically with as fast as 50 Gb/sec rate.
\subsection{Limitations and Analysis}
In this subsection, proposed analog optical computing structure is analyzed in view of repetition rate of the input optical pulses, band-width, the required waveguide length and the frequency-change-rate of the quadratic phase modulation coefficient.\par
The required bandwidth is determined by the range of frequencies where the utilized dispersive media for performing $FT$ and $IFT$ has a linear or constant GVD and also by the optical-modulator band-width. According to Fig. \ref{fig:fig6}, for example the silicon on insulator waveguide of length $L = 6.78~cm$ has a constant GVD of  $\ddot \beta=7.374\times10^{3} ~(\frac{{{ps^2}}}{km})$ in the $193.9$ to $194.4~THz$ frequency range which corresponds to $500~GHz$ bandwidth, appropriate for high-speed optical communication systems. The input pulse duration should be less than the value of the group-delay. Therefore, due to the time-frequency mapping relation presented in Eq. (\ref{eq5}) and given that the pulse-repetition-rate, R, should be less than the inverse of group delay dispersion, $\Delta {\tau ^{ - 1}}$, the $500~GHz$ bandwidth allows us to have the maximum pulse repetition rate of $0.1~THz$ as follows: 
\begin{equation}
\label{eq13}
\begin{array}{l}
R < \Delta {\tau ^{ - 1}}\\
\Rightarrow R < {(\left( {\ddot \beta L} \right)\Delta \omega )^{ - 1}}\\
\Rightarrow R < 0.1THz
\end{array}
\end{equation}
which, implies pulse duration of less than 0.1 $psec$. On the other side, the maximum bandwidth of the existing modulators is potentially more than 50 $GHz$ corresponding to input pulse duration of less than 0.02 $nsec$, which may limit the communication rate. However, as this field is developing, by having modulators of higher bandwidth, we can use the proposed structure for high-speed optical communication.\par
As an important parameter, the length of the dispersive medium should be considered which in an integrated system should be as short as possible. According to Eq. (\ref{eq4}), the input signal is modulated by a quadratic phase factor of $b = \frac{1}{{2{{\ddot \varphi }_0}}}$. Given that ${\ddot \varphi _0} = L\ddot \beta $ , the required length of the dispersive medium is obtained as follows:
\begin{equation}
\label{eq14}
L = \frac{1}{{2b\ddot \beta }}
\end{equation}
value of the group velocity dispersion, $\ddot \beta$, depends on the type of the dispersive medium and varies between ${26.8}~p{s ^2}/{km}$  for single mode fiber (SMF) to ${5}\times{10}^{7}~p{s^2}/{km}$    for a LCFBG as represented by the slope of dispersion curves in Fig. \ref{fig:fig6}. According to LCFBG value and given that for a $T=10~ps$ input pulse the chrip phase modulation factor, $b$, should be larger than ${(\frac{1}{T})^2} = {10^{ - 2}}~p{s ^{ - 2}}$, hence, the length of the waveguide could be chosen to be approximately as small as $1 ~mm$. Therefore the chirp phase modulation factor can be adjusted with respect to GVD in a way to reduce required waveguide length making it practical for on-chip implementation. It is worth to mention that in LCFBG, the refractive index profile of the grating is chirped so that different wavelengths reflected from the grating undergo different time delays, thus leading to a large GVD (Fig. \ref{fig:fig4}). The advantage of the LCFBG for DFT is its short length and ability to customize the amount of GVD easily, while its disadvantage includes group-velocity ripples~\cite{8}. Theses ripples in turn will be noticed in time-frequency curve of the LCFBG as shown in Fig. \ref{fig:fig6}\subref{subfig6:b}. Thus, the effect of these ripples on the output signal is important and leads to a non-ideal Fourier transform. It is also important to note that since the product of the total GDD, $\Delta \tau $,  and the optical bandwidth, $\Delta \omega$, should be less than the repetition period of the laser, for a given laser repetition rate and large bandwidth, there is an upper limit on the amount of GVD \cite{8}.

\section{Simulation Results}
\begin{figure}
	\centering
	\includegraphics[height=9cm,width=1\linewidth]{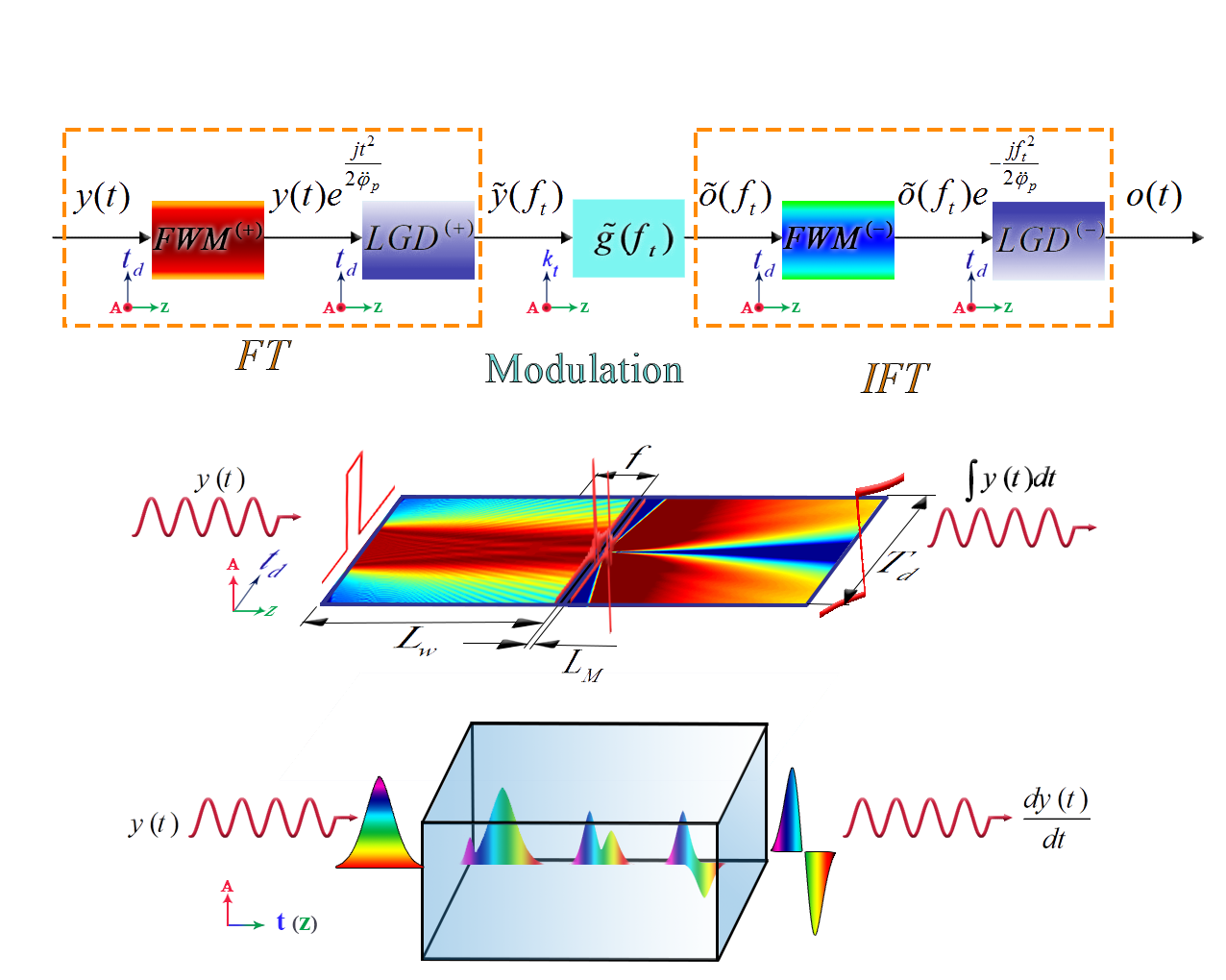}
	\caption{Schematic demonstration of the concept of temporal optical computing using time lens and optical modulator.}
	\label{fig:fig8}
\end{figure}

In this section, different Green's functions corresponding to a differentiator and an integrator, as well as a Green's function of an arbitrary shape, are implemented through the proposed structure. The simulations are performed using commercially available softwares, i.e. Lumerical Mode Solutions and MATLAB. As discussed in the previous section, a linear and positive slope group delay with respect to wavelength, representing as $LGD^{(+)}$, performs Fourier transform on the impinging optical temporal signal. Adversely, in the case of negative slope, the inverse Fourier transform is carried out indicating by $LGD^{(-)}$. In this manner, optical computing is performed using two silicon on insulator dispersive waveguides with  up to second order dispersion coefficients, $L_S$ length, $W_S$ width and $L_w$ long for $FT$ and $IFT$. In order to avoid complexity of the structure $FT$ is obtained instead of $IFT$ with this in mind that $FT\{ \tilde o(f_t)\}  = o(-t)$. An MZI-based optical modulator with $L_M$ long is placed in between for multiplying arbitrary Green's function in the Fourier domain. As the length of $L_M$ is usually comparable to $L_W$, it stands to reason that a corresponding shift be obtained in linearly chirp modulation as discussed in section II. Finally, the proposed optical computing structure is schematically illustrated in Fig. \ref{fig:fig8}.\par
In order to implement the first-order differentiation by the proposed structure, the Green's function produced by the modulator should obey the following relation:
\begin{equation}
\label{eq15}
\tilde g({k_t}) \propto (i{k_t} + i{k_{{t_0}}})
\end{equation}
which, when multiplied by the $FT$ of the input signal resulted from the dispersive waveguide with positive GVD and then passed through the other waveguide with negative GVD for performing $IFT$, yields the first-order differentiation of the input. ${k_{{t_0}}}$ is added to avoid zero multiplication at $t=0$. To implement the n-the order differentiator, we can put these temporal optical computing (TOC) blocks in series or change the Green's function in the modulator as follows:
\begin{equation}
\label{eq16}
\tilde g({k_t}) \propto {(i{k_t} + i{k_{{t_0}}})^n}
\end{equation}
For the integration operation, the Green's function produced by the modulator should be
\begin{equation}
\label{eq17}
\tilde g({k_t}) \propto {(i{k_{{t_0}}} + d/i{k_t})^{ - 1}}
\end{equation}
where, $d$ is the normalization constant which avoid the modulator gain requirement. Accordingly, Eq. (\ref{eq16}) is truncated at $d$ and then is applied to the modulator. Although this truncation slightly distorts the low frequency component of the integrator output, its effect can be compensated to a good extent following the method discussed in \cite{josab}. It is worth noting that the n-th order integrator can also be accomplished using these TOC blocks in series or by change the modulator function to $\tilde g({k_t}) \propto {(d/i{k_t})^{ - n}}$.\par
One of the most important operation in the optical computing regime is convolution~\cite{1,2,3}. For example, an image can be sampled in time and goes through a convolution structure for edge detection~\cite{1}.  Convolution with an arbitrary kernel can be described as:
\begin{equation}
\label{eq18}
\tilde g({k_t}) \propto \Im \{ z(t)\} 
\end{equation}
where, $z(t)$ is an arbitrary kernel.\par
\begin{figure*}
	\centering
	\includegraphics[width=.9\linewidth]{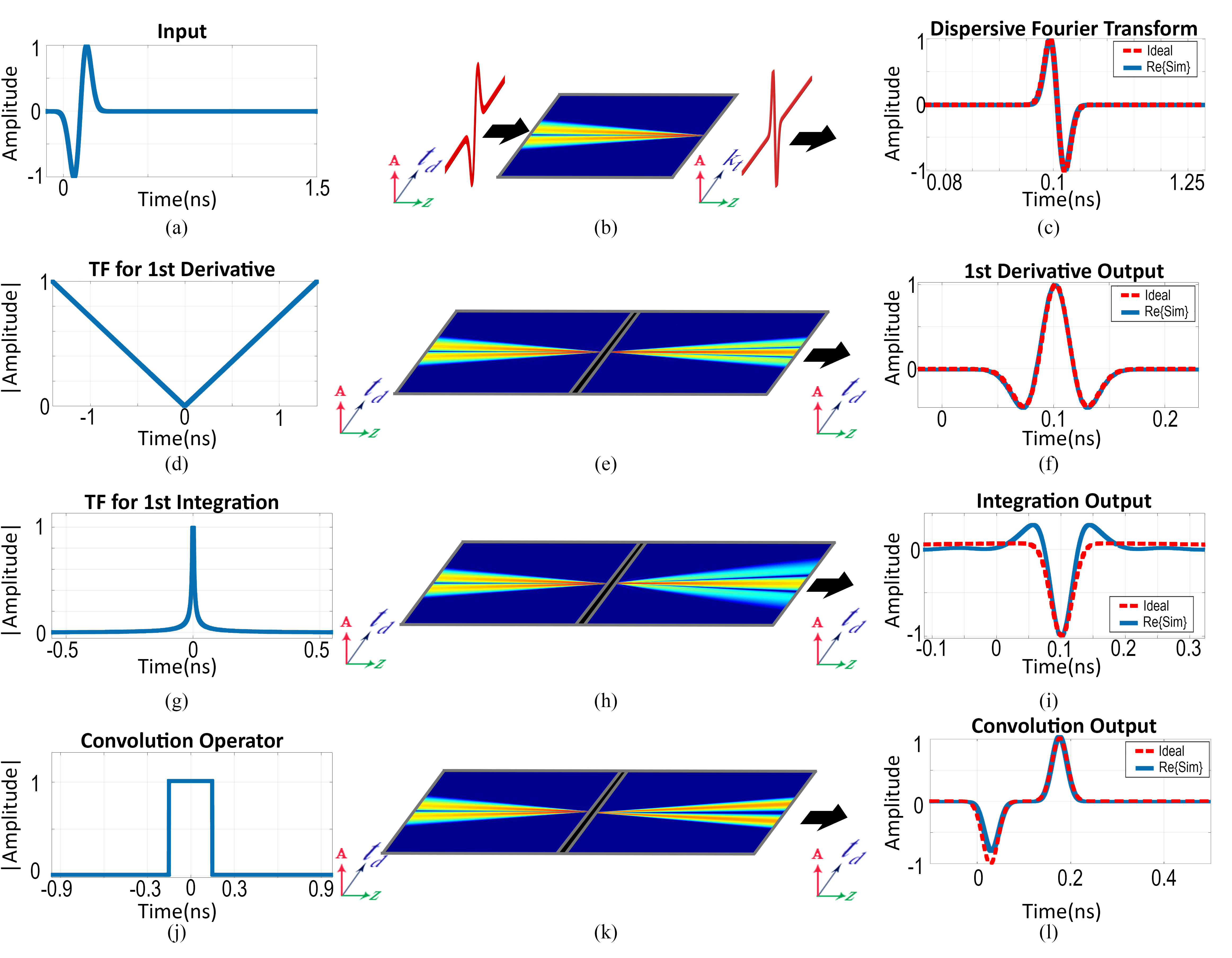}
	\caption{Temporal optical computing using dispersive Fourier transform and optical modulator. (a)Input function $y(t) = t{e^{ - {{(\frac{{2t}}{{{T_p}}})}^2}}}$  propagating through the dispersive medium with up to second order dispersion coefficient in the perspective view of (b). (c)Simulation results for dispersive Fourier transform of y(t). (d)Absolute differentiation TF applied to the modulator. (e)Temporal optical computing (TOC) perspective view of differentiation. (f)Real part of differentiation output using Schr\"{o}dinger equation versus its ideal value. (g)TF for integration. (h)TOC perspective view for integration. (i)Simulation result of the real part of integration output. (j)Convolution kernel applied to modulator. (k)Perspective view for convolving $y(t)$ with a rectangular kernel. (l)Simulation result of the real part of convolution once passed through TOC.}
	\label{fig:fig9}
\end{figure*}

All these Green's functions are simulated and compared with ideal results through Schr\"{o}dinger equation and dispersion parameter of the waveguide obtained from photonic crystal (PC) waveguide presented in Fig. \ref{fig:fig6} \cite{ogawa2006broadband}. The main part of the PC waveguide is low-loss $Si_{3}N_{4}$ rectangular core with $1 \mu m$ width and $400 nm$ thickness. Si/SiO2 photonic crystal layer underlying the core consists of two-dimensional triangular lattice of SiO2 pillars  with pitch of -400nm embedded in host Si layer. The diameter of SiO2 pillar is 250nm. Thin SOI layer (lOOnm thickness) serves as host medium for the PC layer, and buried oxide layer with $1 \mu m$ thickness in SOI wafer as bottom cladding. Moreover, in our simulations, we assumed ${L_{p}}=L_{w}=10.56~mm,~{L_M}=4~mm,~{L_S}=400~nm,~{W_S}=1 ~\mu m,~\dot \beta_p  =\dot \beta_{w}= 3.7735 \times {10^{4}} ~\frac{{{ps}}}{km},~\ddot \beta_p  =\ddot \beta_{w}= 2.81 \times {10^{6}}~\frac{{{ps^2}}}{km},~b = 1.67 \times {10^{-2}}~{ps ^{ - 2}},~d = 1~ ps$ and the input function $y(t) = t{e^{ - {{(\frac{{2t}}{{{T_p}}})}^2}}}$ where, ${T_p} = 0.1~ns$ is the input pulse width. Fig. \ref{fig:fig9} shows the simulation results of the proposed temporal optical computing (TOC) structure. The input, $y(t)$, and its Fourier transform from the dispersive medium are shown in Fig. \ref{fig:fig9}(a) and (c), respectively. The output result for integration,differentiation, and convolution operations are also shown in Fig. \ref{fig:fig9}. The transfer function corresponding to Green's function of each of these operations are shown in Fig. \ref{fig:fig9}(d), (g), and (j), while the outputs of TOC versus their ideal values are plotted in Fig. \ref{fig:fig9}(f,i,l) for integration, differentiation, and convolution operations, respectively. Perspective view of operations are shown by plotting logarithmic amplitude versus delay and distance in Figure \ref{fig:fig9}(b,e,h,k). This perspective view demonstrates the temporal focal length and time to frequency mapping behavior of the input optical pulse propagating through a dispersive medium.
\section{validation and considerations}
\subsection{Distance and Chirp Modulation Duality}
\begin{figure}
	\centering
	\mbox{\subfloat[]{\label{subfig10:a} \includegraphics[height=3.7cm]{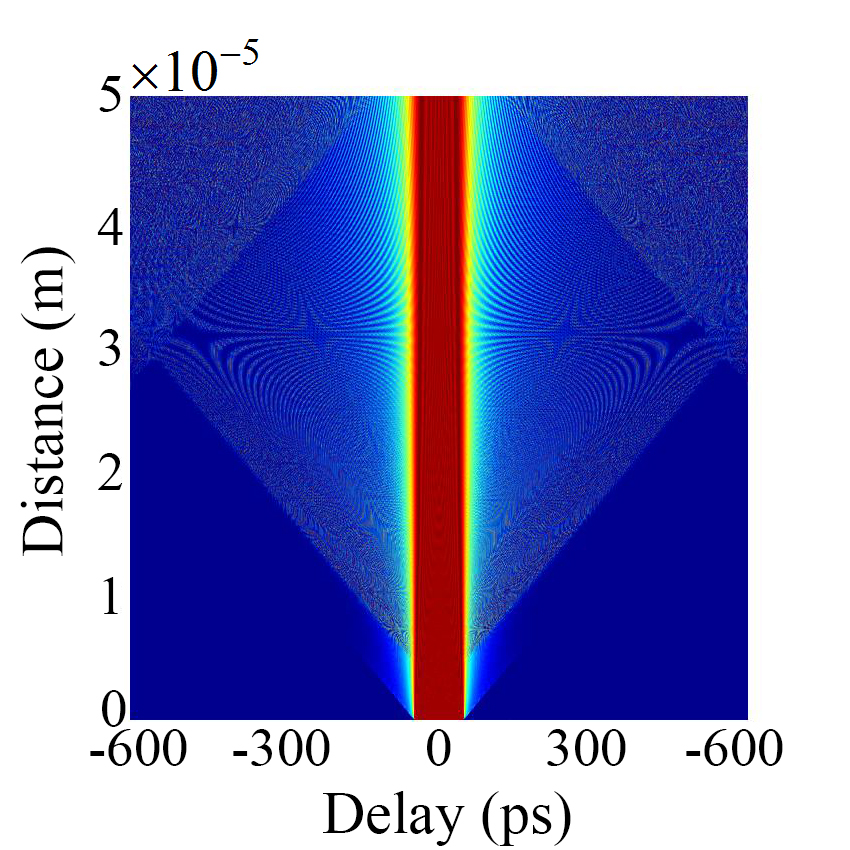}}}
	\mbox{\subfloat[]{\label{subfig10:b} \includegraphics[height=3.7cm]{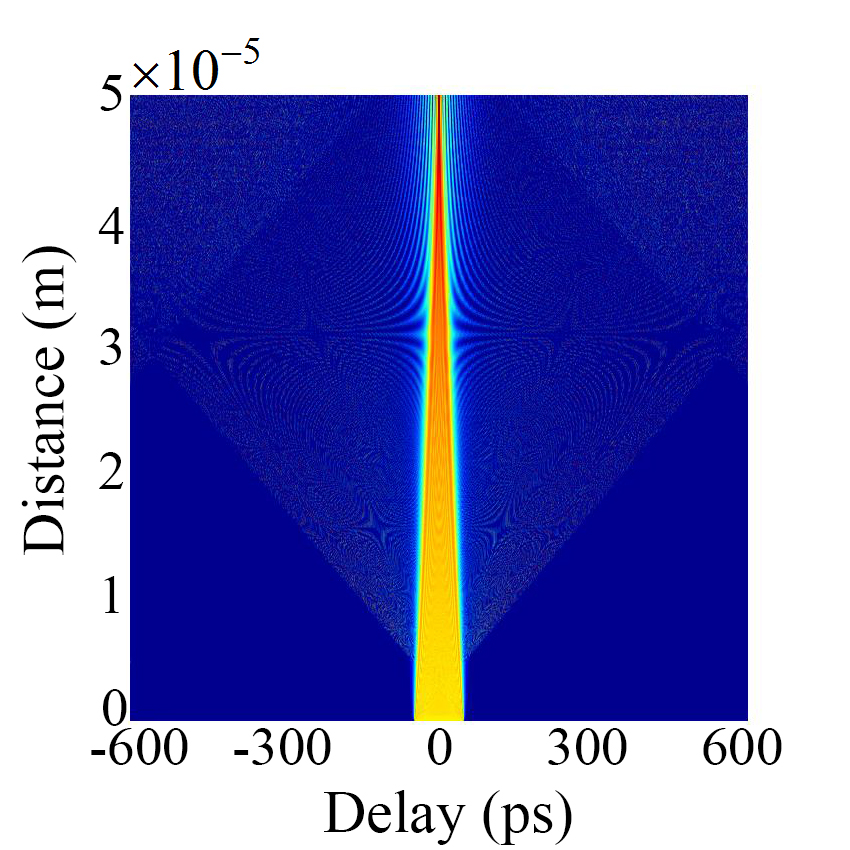}}}\\
	\mbox{\subfloat[]{\label{subfig10:c} \includegraphics[height=3.7cm]{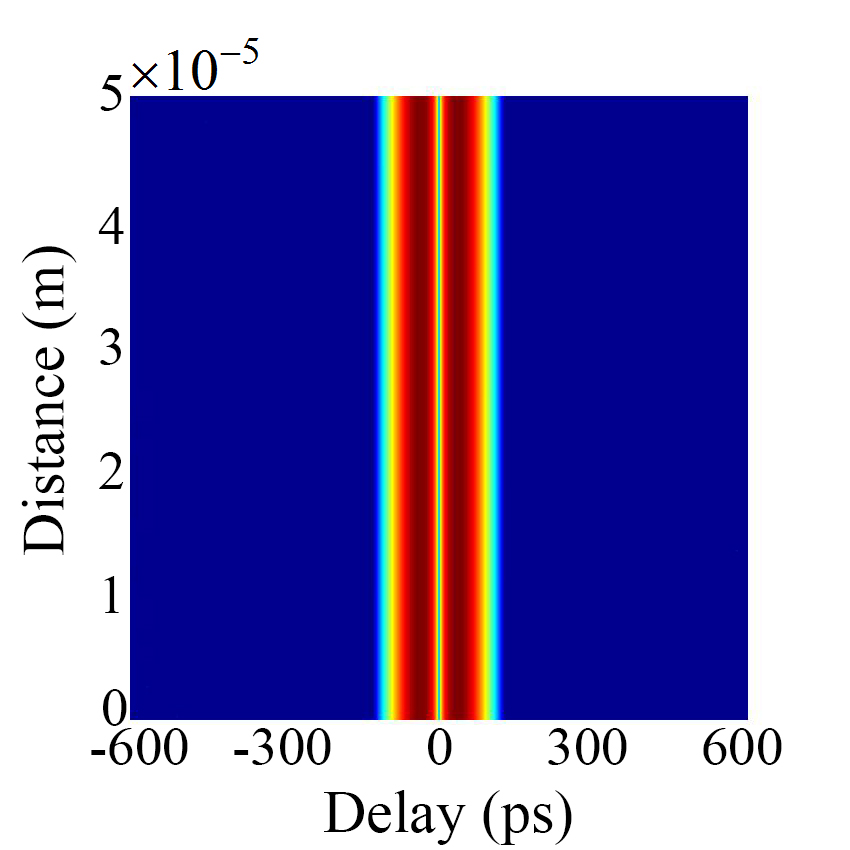}}}
	\mbox{\subfloat[]{\label{subfig10:d} \includegraphics[height=3.7cm]{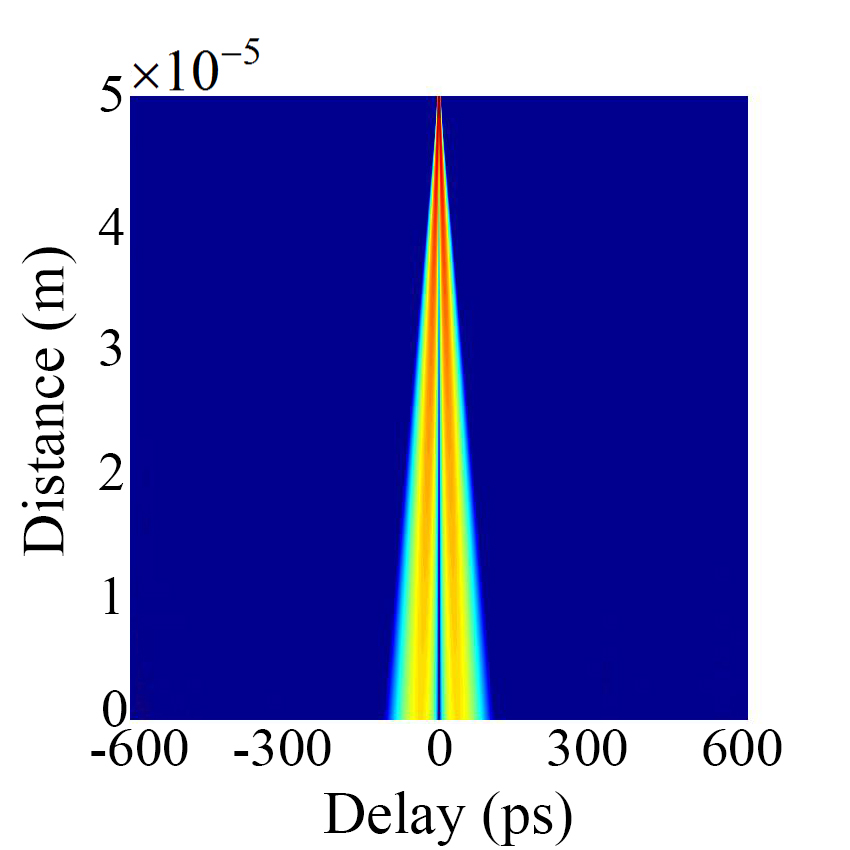}}}
	\caption{The effect of linearly chirp modulation in reducing the focal length. Rect function of with $T_p$ and $y(t) = t{e^{ - {{(\frac{{2t}}{{{T_p}}})}^2}}}$  are propagating as input signal without linearly chirp modulation in (a) and (c) and with linearly chirp modulation in (b) and (d), respectively.}
	\label{fig:fig10}		
\end{figure}

The length of a dispersive waveguide, translated to its focal length in DFT usage, can be approximated as follows~\cite{10}:
\begin{equation}
\label{eq19}
f = \frac{1}{{2\ddot \beta b}}
\end{equation}
where, $b$ is the chirpyness coefficient in the modulation chirp function applied to the input signal represented in Eq. (\ref{eq3}) as  ${e^{ - j\frac{1}{{2{\beta _2}z}}{\tau _R}^2}}$. The following equation can be derived by replacing the chirpyness coefficient, $b = \frac{1}{{2{{\ddot \beta }_p}{L_p}}}$, from Eq. (\ref{eq4}): 
\begin{equation}
\label{eq20}
f = {L_p}\frac{{{{\ddot \beta }_p}}}{{{{\ddot \beta }_w}}} = {L_{w}}
\end{equation}
As discussed in section 3.C, given that $\ddot \beta_w$ has a relatively small value, a long waveguide is required to implement Fourier transformation. Still, increasing the value of $b$ helps decreasing the waveguide length  to practical values required for photonic integrated circuit implementation. The limiting factor is the input pulse width which imposes a lower limit on the value of chirp coefficient, $b$. In fact, the chirpyness factor, $b$, should be large enough to guarantee that enough number of different frequencies, as well as, enough number of periods in each frequency fall in the pulse duration time. As an example, for our waveguide with $\ddot \beta$  equal to ${1.043\times 10^9}~\frac{{p{{s }^2}}}{km}$, a pulse width of $100~ps$ imposes the lower limit of ${(\frac{1}{T})^2} = {10^{ - 4}}~p{s ^{ - 2}}$, on chirp factor, $b$, which in turn limits the length to the maximum value of $5~mm$. Fig. \ref{fig:fig10} demonstrates the effect of chirp modulation in reducing the focal length for $b \approx 100{b_{\min }}$ and performing dispersive Fourier transformation with a waveguide length as short as $50~\mu m$. This limitation can be formolized by using the fact that $b > {b_{\min }}$ which yields ${L_p} < \frac{T}{{20{{\ddot \beta }_p}}}$.

Another important factor that should be considered is the operation length which applies the lower limit on the waveguide length. The operation length is defined as the length of the generated chirp pulse ready for multiplication which should be larger the maximum required length for computation.This limitation means that ${\tau _{operation}} > {\tau _{\max }}$, which coresponds to $\frac{{{\tau _{min}}}}{{2{{\ddot \beta }_p}{\Omega _p}}} < {L_p}$. Therefore, considering the fact that ${L_w} = {L_p}\frac{{{{\ddot \beta }_p}}}{{{{\ddot \beta }_w}}}$, the waveguide length should obey the following equation:
\begin{equation}
\label{eq21}
\frac{{{\tau _{max}}}}{{2{{\ddot \beta }_p}{\Omega _p}}} < {L_p} < \frac{T}{{20{{\ddot \beta }_p}}}
\end{equation}
where $\Omega _p$ and $T$ are the spectral bandwidth of the pump pulse and the input pulse width, respectively.

In order to calculate the minimum wavelength for the signal and pump waveguide, all of the aforementioned equations should be considered. The minimum length for the waveguide enforce that ${L_w} = {L_p}$ resulting in ${{\ddot \beta }_p} = {{\ddot \beta }_{w}}$. 

\subsection{Linear Schr\"{o}dinger Equation}
Linear Schr\"{o}dinger equation is a classical field equation which describes temporal behavior of light in optical fibers or planar waveguide with no nonlinear property \cite{44}. Classical form of this equation can be derived from Maxwell’s equation with some simplifying assumptions, such as having no charge in the structure and zero current density, as follows~\cite{48}:
\begin{equation}
\label{eq22}
\begin{array}{l}
\frac{{\partial A}}{{\partial z}} + \frac{\alpha }{2}A + \sum\limits_{n = 2}^k {\frac{{{i^{n - 1}}{\beta _n}}}{{n!}}\frac{{{\partial ^n}A}}{{\partial {t^n}}}}  =0
\end{array}
\end{equation}
where, $A$ is the pulse envelope, ${\beta _n}$ is the n-the order dispersion coefficient, and $\alpha$ indicates waveguide loss in $dB/Km$. To validate the proposed computational approach, in the simulation framework, we used linear Schr\"{o}dinger equation as the golden metric to verify the approximation results from LTI system and validity of the proposed TOC structure. As shown in Fig. \ref{fig:fig9}, this comparison demonstrates a high accuracy of the presented results. In the Schr\"{o}dinger equation simulation, we assumed $\alpha  = 0.8~dB/m$, $\beta_1  = 1.4 \times {10^{7}} ~\frac{{{ps}}}{km}$ and $\beta_2  = 1.043 \times {10^{9}}~\frac{{{ps^2}}}{km}$.

\section{Conclusion}
We have designed and demonstrated a fully reconfigurable photonic integrated signal processor based on a dispersive Fourier transformer and time lens. The optical temporal signal processor can be reconfigured as a photonic temporal integrator, differentiator, and convolution operator, as the basic building blocks required for general-purpose signal processing. The proposed photonic signal processor can be used to perform any arbitrary mathematical operation which can be modeled by a transfer function. The in-depth analysis of the work was also carried out revealing $100 ps$ operation time and a high resolution of $300 fs$. In particular, our work suggests a fully reconfigurable high speed general-purpose photonic signal processor overcoming the inherent speed limitation of electronic signal processors.

\section*{References}

\bibliography{References}

\begin{thebibliography}{10}
\expandafter\ifx\csname url\endcsname\relax
  \def\url#1{\texttt{#1}}\fi
\expandafter\ifx\csname urlprefix\endcsname\relax\def\urlprefix{URL }\fi
\expandafter\ifx\csname href\endcsname\relax
  \def\href#1#2{#2} \def\path#1{#1}\fi

\bibitem{1}
W.~Liu, M.~Li, R.~S. Guzzon, E.~J. Norberg, J.~S. Parker, M.~Lu, L.~A. Coldren,
  J.~Yao, A fully reconfigurable photonic integrated signal processor, Nature
  Photonics 10~(3) (2016) 190--195.

\bibitem{2}
R.~Nogueira, R.~Oliveira, L.~Bilro, J.~Heidarialamdarloo, New advances in
  polymer fiber bragg gratings, Optics \& Laser Technology 78 (2016) 104--109.

\bibitem{3}
F.~Garrelie, F.~Bourquard, C.~Donnet, J.-P. Colombier, et~al., Control of
  femtosecond pulsed laser ablation and deposition by temporal pulse shaping,
  Optics \& Laser Technology 78 (2016) 42--51.

\bibitem{4}
T.~L. Huang, A.~L. Zheng, J.~J. Dong, D.~S. Gao, X.~L. Zhang,
  Terahertz-bandwidth photonic temporal differentiator based on a
  silicon-on-isolator directional coupler, Optics letters 40~(23) (2015)
  5614--5617.

\bibitem{5}
M.~Ferrera, Y.~Park, L.~Razzari, B.~E. Little, S.~T. Chu, R.~Morandotti,
  D.~Moss, J.~Aza{\~n}a, On-chip cmos-compatible all-optical integrator, Nature
  Communications 1 (2010) 29.

\bibitem{6}
T.~Yang, J.~Dong, L.~Lu, L.~Zhou, A.~Zheng, X.~Zhang, J.~Chen, All-optical
  differential equation solver with constant-coefficient tunable based on a
  single microring resonator, Scientific reports 4 (2014) 5581.

\bibitem{7}
A.~M. Weiner, Ultrafast optical pulse shaping: A tutorial review, Optics
  Communications 284~(15) (2011) 3669--3692.

\bibitem{13}
W.~Zhao, S.~Liu, H.~Qi, G.~Peng, M.~Shen, Sampled fiber grating for wdm signal
  queuing with picosecond time interval, Optics \& Laser Technology 97 (2017)
  302--307.

\bibitem{14}
H.~Zhang, S.~Zhao, G.~Li, K.~Yang, D.~Li, Single mode-locking pulse generation
  underneath the q-switched envelope of the doubly qml green laser with eo and
  gaas, Optics \& Laser Technology 45 (2013) 118--124.

\bibitem{15}
A.~E. Willner, S.~Khaleghi, M.~R. Chitgarha, O.~F. Yilmaz, All-optical signal
  processing, Journal of Lightwave Technology 32~(4) (2014) 660--680.

\bibitem{16}
C.~Koos, P.~Vorreau, T.~Vallaitis, P.~Dumon, W.~Bogaerts, R.~Baets,
  B.~Esembeson, I.~Biaggio, T.~Michinobu, F.~Diederich, et~al., All-optical
  high-speed signal processing with silicon--organic hybrid slot waveguides,
  Nature photonics 3~(4) (2009) 216--219.

\bibitem{18}
V.~R. Almeida, C.~A. Barrios, R.~R. Panepucci, M.~Lipson, All-optical control
  of light on a silicon chip, Nature 431~(7012) (2004) 1081--1084.

\bibitem{19}
C.~R. Doerr, K.~Okamoto, Advances in silica planar lightwave circuits, Journal
  of lightwave technology 24~(12) (2006) 4763--4789.

\bibitem{20}
R.~Slav{\'\i}k, Y.~Park, N.~Ayotte, S.~Doucet, T.-J. Ahn, S.~LaRochelle,
  J.~Aza{\~n}a, Photonic temporal integrator for all-optical computing, Optics
  express 16~(22) (2008) 18202--18214.

\bibitem{21}
M.~Michalska, J.~Swiderski, M.~Mamajek, Arbitrary pulse shaping in er-doped
  fiber amplifiers—possibilities and limitations, Optics \& Laser Technology
  60 (2014) 8--13.

\bibitem{22}
A.~Jolly, P.~Estraillier, Generation of arbitrary waveforms with electro-optic
  pulse-shapers for high energy—multimode lasers, Optics \& Laser Technology
  36~(1) (2004) 75--80.

\bibitem{azana1}
S.~Thomas, A.~Malacarne, F.~Fresi, L.~Pot{\`\i}, A.~Bogoni, J.~Aza{\~n}a,
  Programmable fiber-based picosecond optical pulse shaper using time-domain
  binary phase-only linear filtering, Optics letters 34~(4) (2009) 545--547.

\bibitem{azana2}
S.~Thomas, A.~Malacarne, F.~Fresi, L.~Pot{\`\i}, J.~Aza{\~n}a, Fiber-based
  programmable picosecond optical pulse shaper, Journal of Lightwave Technology
  28~(12) (2010) 1832--1843.

\bibitem{53}
J.~Capmany, D.~Novak, Microwave photonics combines two worlds, Nature photonics
  1~(6) (2007) 319--330.

\bibitem{23}
R.~Ashrafi, M.~R. Dizaji, L.~R. Cort{\'e}s, J.~Zhang, J.~Yao, J.~Aza{\~n}a,
  L.~R. Chen, Time-delay to intensity mapping based on a second-order optical
  integrator: application to optical arbitrary waveform generation, Optics
  express 23~(12) (2015) 16209--16223.

\bibitem{24}
N.~Q. Ngo, et~al., Optical realization of newton-cotes-based integrators for
  dark soliton generation, Journal of lightwave technology 24~(1) (2006) 563.

\bibitem{25}
M.~T. Hill, H.~J. Dorren, T.~De~Vries, X.~J. Leijtens, J.~H. Den~Besten,
  B.~Smalbrugge, Y.-S. Oei, H.~Binsma, G.-D. Khoe, M.~K. Smit, A fast low-power
  optical memory based on coupled micro-ring lasers, nature 432~(7014) (2004)
  206--209.

\bibitem{26}
E.~Reeves, P.~Costanzo-Caso, A.~Siahmakoun, Theoretical study and demonstration
  of photonic asynchronous first-order delta-sigma modulator for converting
  analog input to nrz binary output, Microwave and Optical Technology Letters
  57~(3) (2015) 574--578.

\bibitem{27}
M.~Ferrera, Y.~Park, L.~Razzari, B.~E. Little, S.~T. Chu, R.~Morandotti,
  D.~Moss, J.~Aza{\~n}a, On-chip cmos-compatible all-optical integrator, Nature
  Communications 1 (2010) 29.

\bibitem{28}
M.~Ferrera, Y.~Park, L.~Razzari, B.~E. Little, S.~T. Chu, R.~Morandotti, D.~J.
  Moss, J.~Aza{\~n}a, All-optical 1st and 2nd order integration on a chip,
  Optics express 19~(23) (2011) 23153--23161.

\bibitem{29}
F.~Liu, T.~Wang, L.~Qiang, T.~Ye, Z.~Zhang, M.~Qiu, Y.~Su, Compact optical
  temporal differentiator based on silicon microring resonator, Optics Express
  16~(20) (2008) 15880--15886.

\bibitem{30}
D.~Hillerkuss, M.~Winter, M.~Teschke, A.~Marculescu, J.~Li, G.~Sigurdsson,
  K.~Worms, S.~B. Ezra, N.~Narkiss, W.~Freude, et~al., Simple all-optical fft
  scheme enabling tbit/s real-time signal processing, Optics express 18~(9)
  (2010) 9324--9340.

\bibitem{31}
D.~Hillerkuss, R.~Schmogrow, T.~Schellinger, M.~Jordan, M.~Winter, G.~Huber,
  T.~Vallaitis, R.~Bonk, P.~Kleinow, F.~Frey, et~al., 26 tbit s-1 line-rate
  super-channel transmission utilizing all-optical fast fourier transform
  processing, Nature Photonics 5~(6) (2011) 364--371.

\bibitem{32}
F.~Li, Y.~Park, J.~Aza{\~n}a, Complete temporal pulse characterization based on
  phase reconstruction using optical ultrafast differentiation (proud), Optics
  letters 32~(22) (2007) 3364--3366.

\bibitem{33}
R.~Slavik, L.~Oxenlowe, M.~Galili, H.~C.~H. Mulvad, Y.~Park, J.~Aza{\~n}a,
  P.~Jeppesen, Demultiplexing of 320-gb/s otdm data using ultrashort flat-top
  pulses, IEEE Photonics Technology Letters 19~(22) (2007) 1855--1857.

\bibitem{34}
Y.~Xu, M.~Savescu, K.~R. Khan, M.~F. Mahmood, A.~Biswas, M.~Belic, Soliton
  propagation through nanoscale waveguides in optical metamaterials, Optics \&
  Laser Technology 77 (2016) 177--186.

\bibitem{51}
A.~Silva, F.~Monticone, G.~Castaldi, V.~Galdi, A.~Al{\`u}, N.~Engheta,
  Performing mathematical operations with metamaterials, Science 343~(6167)
  (2014) 160--163.

\bibitem{52}
S.~AbdollahRamezani, K.~Arik, A.~Khavasi, Z.~Kavehvash, Analog computing using
  graphene-based metalines, Optics letters 40~(22) (2015) 5239--5242.

\bibitem{josab}
H.~Babashah, Z.~Kavehvash, S.~Koohi, A.~Khavasi, Integration in analog optical
  computing using metasurfaces revisited: toward ideal optical integration,
  JOSA B 34~(6) (2017) 1270--1279.

\bibitem{8}
K.~Goda, B.~Jalali, Dispersive fourier transformation for fast continuous
  single-shot measurements, Nature Photonics 7~(2) (2013) 102--112.

\bibitem{9}
C.~Wang, Dispersive fourier transformation for versatile microwave photonics
  applications, in: Photonics, Vol.~1, Multidisciplinary Digital Publishing
  Institute, 2014, pp. 586--612.

\bibitem{10}
T.~Jannson, Real-time fourier transformation in dispersive optical fibers,
  Optics letters 8~(4) (1983) 232--234.

\bibitem{foster2008silicon}
M.~A. Foster, R.~Salem, D.~F. Geraghty, A.~C. Turner-Foster, M.~Lipson, A.~L.
  Gaeta, Silicon-chip-based ultrafast optical oscilloscope, Nature 456~(7218)
  (2008) 81.

\bibitem{11}
M.~Liu, X.~Yin, E.~Ulin-Avila, B.~Geng, T.~Zentgraf, L.~Ju, F.~Wang, X.~Zhang,
  A graphene-based broadband optical modulator, Nature 474~(7349) (2011)
  64--67.

\bibitem{111}
C.~T. Phare, Y.-H.~D. Lee, J.~Cardenas, M.~Lipson, Graphene electro-optic
  modulator with 30 ghz bandwidth, Nature Photonics 9~(8) (2015) 511--514.

\bibitem{12}
M.~A. Muriel, J.~Aza{\~n}a, A.~Carballar, Real-time fourier transformer based
  on fiber gratings, Optics letters 24~(1) (1999) 1--3.

\bibitem{Hydex}
D.~J. Moss, R.~Morandotti, A.~L. Gaeta, M.~Lipson, New cmos-compatible
  platforms based on silicon nitride and hydex for nonlinear optics, Nature
  Photonics 7~(8) (2013) 597--607.

\bibitem{45}
D.~Tan, K.~Ikeda, P.~Sun, Y.~Fainman, Group velocity dispersion and self phase
  modulation in silicon nitride waveguides, Applied Physics Letters 96~(6)
  (2010) 061101.

\bibitem{46}
M.~A. Foster, A.~C. Turner, J.~E. Sharping, B.~S. Schmidt, M.~Lipson, A.~L.
  Gaeta, Broad-band optical parametric gain on a silicon photonic chip, Nature
  441~(7096) (2006) 960--963.

\bibitem{37}
X.~Gao, J.~Yuan, Y.~Yang, Y.~Li, W.~Cai, X.~Li, Y.~Wang, Light coupling for
  on-chip optical interconnects, Optics \& Laser Technology 97 (2017) 154--160.

\bibitem{38}
J.~Cardenas, C.~B. Poitras, K.~Luke, L.-W. Luo, P.~A. Morton, M.~Lipson, High
  coupling efficiency etched facet tapers in silicon waveguides, IEEE Photon.
  Technol. Lett. 26~(23) (2014) 2380--2382.

\bibitem{ogawa2006broadband}
K.~Ogawa, K.~Tomiyama, Y.~T. Tan, M.~T. Doan, Y.~M. Bin, D.-L. Kwong,
  S.~Yamada, J.~Cole, Y.~Katayama, H.~Mizuta, et~al., Broadband variable
  chromatic dispersion in photonic-band electro-optic waveguide, in: Optical
  Fiber Communication Conference, 2006 and the 2006 National Fiber Optic
  Engineers Conference. OFC 2006, IEEE, 2006, pp. 3--pp.

\bibitem{35}
{\'A}.~Gonz{\'a}lez-Vila, D.~Kinet, P.~M{\'e}gret, C.~Caucheteur, Narrowband
  interrogation of plasmonic optical fiber biosensors based on spectral combs,
  Optics \& Laser Technology 96 (2017) 141--146.

\bibitem{36}
G.~T. Reed, G.~Mashanovich, F.~Gardes, D.~Thomson, Silicon optical modulators,
  Nature photonics 4~(8) (2010) 518--526.

\bibitem{44}
X.~Tu, T.-Y. Liow, J.~Song, X.~Luo, Q.~Fang, M.~Yu, G.-Q. Lo, 50-gb/s silicon
  optical modulator with traveling-wave electrodes, Optics express 21~(10)
  (2013) 12776--12782.

\bibitem{48}
G.~P. Agrawal, Nonlinear fiber optics, Academic press, 2007.

\end{thebibliography}

\end{document}